\documentclass[aps,twocolumn,showpacs,prl]{revtex4}
\usepackage{amssymb}
\usepackage{natbib}
\usepackage{amsmath}
\usepackage{amsfonts}
\usepackage{graphicx}
\usepackage{mathrsfs}
\usepackage{dcolumn}
\usepackage{bm}
\usepackage{mathbbol}
\usepackage{color}
\definecolor{rot}{rgb}{0.75,0.05,0.25}
\definecolor{hellgrau}{gray}{0.5}
\definecolor{blau}{rgb}{0,0,0.7}

\definecolor{rot}{rgb}{0.75,0.05,0.25}
\definecolor{hellgrau}{gray}{0.5}
\definecolor{blau}{rgb}{0,0,0.7}

\begin{document}

\title{Fluctuation theorems for continuously monitored quantum
fluxes}
\author{Michele Campisi}
\email[]{michele.campisi@physik.uni-augsburg.de}
\author{Peter Talkner}
\email[]{peter.talkner@physik.uni-augsburg.de}
\author{Peter H\"anggi}
\email[]{hanggi@physik.uni-augsburg.de}
\address{Institute of Physics, University of Augsburg,
Universit\"atsstrasse 1, D-86153 Augsburg, Germany}
\date{\today}

\begin{abstract}
It is shown that quantum fluctuation theorems remain unaffected
if measurements of any kind and number of observables are
performed during the action of a force protocol. 
That is, although the backward and forward probabilities entering
the fluctuation theorems 
are both altered by these measurements, their ratio remains
unchanged.
This observation allows to describe the measurement of fluxes through interfaces and, in this way, to bridge
the gap between the current theory,  based on only two
measurements performed at the beginning and end of the protocol, and experiments that are based on continuous monitoring.
\end{abstract}

\pacs{
05.70.Ln, 
05.30.-d,  
73.63.-b, 
structures
03.65.Ta  
}
\keywords{}

\maketitle
In the last decade fluctuation theorems have experienced a
renewed and widespread interest
\cite{BochkovKuzovlev77,Evans_PRL93,Gallavotti_PRL95,Crooks_PRE99,Esposito_RMP09,HanggiPRL10}.
These theorems yield rigorous
predictions for nonequilibrium processes beyond linear response
theory. In particular, they quantify the probability of events
that are forbidden by the second law of thermodynamics as being
exponentially suppressed compared to the probability of a
typical, allowed event.

While experimental verifications of the fluctuation theorems for
classical systems were performed by different groups and for
different systems, e.g., 
\cite{Collin_Nature05,Liphardt_Science02,Douarche_EPL05}, an
experiment with quantum devices was performed only very recently
\cite{UtsumiPRB10}.
In that experiment the flow of electrons
through a double quantum dot placed between two leads with
different chemical potentials was continuously monitored and the
quantum fluctuation theorem  was verified for the probability
$p(q)$ that a number $q$ of electrons is exchanged by the leads
in a certain interval of time $\tau=t_f-t_0$, see Fig. 1.
\begin{figure}[t]
\includegraphics[width=5cm]{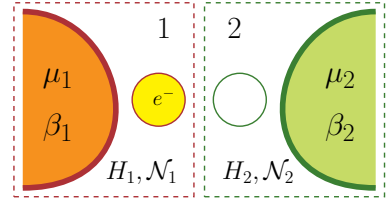}
\caption{(Color online) Schematic representation of a
bidirectional counting statistics experiment. Two leads (large
semicircles) with different electronic chemical potentials
($\mu_1\neq\mu_2$) and same temperature ($\beta_1=\beta_2$) are
connected through a double quantum dot (small circles), whose
quantum state is continuously monitored. The state (1,0), i.e.,
``one electron in the left dot, no electrons in the right dot''
is depicted. The transition from this state to the state (0,1)
signals the exchange of one electron from subsystem 1 to
subsystem 2. $H_{1,2}$ and $\mathcal{N}_{1,2}$ denote the
Hamiltonian and number of electrons operators of each subsystem,
respectively.}
\label{fig:fig1}
\end{figure}
With regard to theory, however, the existing derivations of
fluctuation theorems for quantum transport do not allow for a continuous
monitoring of the flux of energy and/or particles. They are
rather based on two projective quantum measurements of the
relevant observables (Hamiltonian, number of particles),
performed at $t_0$ and $t_f$, where the exchanged energy and
number of particles are obtained as differences of the outcomes
of these two measurements
\cite{JarzPRL04,Saito_PRB08,Andrieux_NJP08,Esposito_RMP09}. 
Evidently, this scheme presents practical difficulties, as
it is not clear, for example, how one could measure the
macroscopically large number of electrons in a lead with the 
required single electron resolution.
Therefore, as other authors also have pointed out (see the
conclusions in Ref. \cite{Andrieux_NJP08}), it is necessary to
bridge the gap between theory and experiment by extending
the theory in such a way as to account for the possibility of
continuously monitoring a specific  quantum observable of the
system.

In this Letter we develop a multiple measurement approach to quantum 
fluctuation theorems that extends the previous two-measurement 
fluctuation theorems \cite{JarzPRL04,Saito_PRB08,Andrieux_NJP08,Esposito_RMP09}
allowing for the possibility to perform measurements of any observable
within the interval $(t_0,t_f)$. 
Most importantly, we demonstrate how to use this approach to overcome
the major problem of
measuring total energies and particle numbers of large reservoirs
 in case of a transport problem as the one
described in Ref. \cite{UtsumiPRB10}. The salient point here is to
consider the \emph{fluxes} through an interface rather than absolute
numbers of particles and amounts of energy. 
This provides a coherent and
general theoretical framework within which the recently reported
electron transport experiment of Ref. \cite{UtsumiPRB10} can be
properly analyzed and understood, and new experiments can
possibly be devised.

\emph{Multiple measurements scheme.$-$}
We consider a quantum system composed of subsystems, whose mutual interaction is turned on only within the
finite time interval $(t_0,t_f)$. They are initially disconnected and separately in equilibrium at different temperatures and chemical potentials. On coupling them through a time-dependent interaction leads to energy and particle transfers. While this exchange of energy and particles occurs a quantum observable is  monitored, as in the experiment reported in
\cite{UtsumiPRB10}, see Fig. 1. 

We begin by
considering only one intermediate measurement occurring at $t_1
\in (t_0,t_f)$. The total Hamiltonian $H(t)$ is: $H(t)=\sum_i
H_i$ for $t \notin (t_0,t_f)$, and $H(t)=\sum_i H_i +V(t)$ for $t
\in (t_0,t_f)$, where $V(t)$ is a time dependent interaction term
that couples the otherwise uncoupled subparts of the system
within the time interval $(t_0,t_f)$.
Due to the time dependence of the Hamiltonian, the microreversibility
principle does no longer apply in the way it does for autonomous systems.
Under the assumption, however, that  \emph{at each instant of time} $t$
the total Hamiltonian commutes with the antiunitary
time-reversal operator $\Theta$, $[H(t),\Theta]=0$, a generalized form of microreversibility is obtained which jointly uses the operator $\Theta$  and the temporal inversion of the protocol $H(t)$ \cite{Andrieux_NJP08,Andrieux08PRL100,CTH-arXiv10} (see the description of the reversed protocol below).

For $t\leq t_0$ each
subsystem is assumed to be at equilibrium with an inverse
temperature $\beta_i$ and chemical potential $\mu_i$ of a
particle species that may be exchanged between the subparts in
presence of interaction. That is, at  $t\leq t_0$ the density
matrix is given by the direct product of two grand
canonical states: $\rho=\prod_i e^{-\beta_i[H_i-\mu_i
\mathcal{N}_i -\Phi_i]}$, with $\Phi_i$ the grand canonical free
energy of subsystem $i$, and $\mathcal{N}_i$ the number operators
of those particles that can be exchanged between the subsystems
once the interaction is turned on. In absence of interaction the
particle numbers in each subsystem are assumed to be conserved,
i.e., $[H_i,\mathcal{N}_i]=0$. In presence of interaction the
total particle number $\mathcal{N}=\sum_i \mathcal{N}_i$ is still
assumed to be conserved, i.e., $[H(t),\mathcal{N}]=0$. 

At time
$t_0$ a joint quantum measurement of all $H_i$'s and
$\mathcal{N}_i$'s is performed. As a consequence the total
system's wave function collapses onto a common eigenstate
$|\psi_n\rangle$ of all these commuting observables, whose
eigenvalues are given by: $H_i|\psi_n\rangle = E^i_n
|\psi_n\rangle$ and $\mathcal{N}_i |\psi_n\rangle = N^i_n
|\psi_n\rangle $. The system then evolves according to the
unitary
time evolution $U_{t_1,t_0}$ governed by $H(t)$ until time $t_1$
when an observable $A$ is measured. We assume for simplicity that
the eigenvalues $a_r$ of $A$ are nondegenerate and that $A$
commutes with 
the time reversal operator $\Theta$. As a
consequence of this measurement, the system's wave function
instantaneously collapses onto an eigenstate $|a_r\rangle$ of
$A$. The evolution generated by $H(t)$ then continues until the
time $t_f$, when a measurement of all $H_i$'s and
$\mathcal{N}_i$'s is again performed, collapsing the wave
function to a joint eigenstate $|\psi_m\rangle$ of $H_i$ and
$\mathcal{N}_i$ with eigenvalues $E^i_m$ and $N^i_m$,
respectively. Accordingly, the changes of energy $\Delta E_i =
E^i_m- E^i_n$ and of particle numbers $\Delta N_i= N^i_m-N^i_n$
are recorded. During each realization of this protocol a
three-state transition of the type $|\psi_n\rangle \rightarrow
|a_r\rangle \rightarrow |\psi_m\rangle$ is recorded. This
three-state transition occurs with probability:
\begin{equation}
P[n,r,m]=p(m,t_f|r,t_1)p(r,t_1|n,t_0) \rho_{nn}
\end{equation}
where
\begin{equation}
\rho_{nn}=\langle \psi_n|\rho
|\psi_n\rangle=\prod_i e^{-\beta_i(
E^i_n-\mu_i N^i_n-\Phi_i)}
\label{eq:rho-nn}
\end{equation}
is the probability to find the system in the state
$|\psi_n\rangle$ at $t_0$, and $p(r,t_1|n,t_0)$
($p(m,t_f|r,t_1)$)
is the conditional probability to find the system in
$|a_r\rangle$ at $t_1$  ($|\psi_m\rangle$ at $t_f$) 
provided that it was in
$|\psi_n\rangle$ at $t_0$ ($|a_r\rangle$ at $t_1$), i.e.:
\begin{align}
p(m,t_f|r,t_1)&=|\langle \psi_m|U_{t_f,t_1}
| a_r\rangle |^2\\
p(r,t_1|n,t_0)&=|\langle a_r|U_{t_1,t_0}
|\psi_n\rangle |^2
\end{align}
Accordingly, the joint forward (F) pdf of energy and particle
number
exchanges with one interruption at $t_1$ becomes:
\begin{align} 
P_F^{t_1}(\{ \Delta E_i\},\{\Delta N_i\})=\sum_{m,n} \prod_i
\delta(\Delta E_i-  E^i_m+ E^i_n)
\nonumber \\
\times \delta(\Delta
N_i- N^i_m+ N^i_n)
p^{t_1}(m,t_f|n,t_0)\rho_{nn}
\label{eq:PFt1}
\end{align}
where,
 \begin{equation}
p^{t_1}(m,t_f|n,t_0) = \sum_r p(m,t_f|r,t_1)p(r,t_1|n,t_0)
\label{eq:chainRule}
\end{equation}
denotes the
conditional probability of finding the state
$|\psi_m\rangle$ at $t_f$ provided that the state $|\psi_n\rangle
$ was found at $t_0$,  and the observable $A$ was measured at
$t_1$. The sum in (\ref{eq:chainRule}) runs over all eigenstates
of $A$, and the superscript $t_1$ in Eqs.
(\ref{eq:PFt1},\ref{eq:chainRule}) indicates the measurement
occurring at $t_1$.

We now consider the reversed protocol specified by preparing
the system in the state $\rho$  introduced above, changing the
Hamiltonian in time according to the reversed schedule
$\widetilde H(t)=H(t_f+t_0-t)$, and performing measurements of
a) all $H_i$'s and $\mathcal{N}_i$'s, at time $t_0$, b) the
observable $A$ at $\widetilde t_1=t_0+t_f-t_1$, c) all $H_i$'s,
$\mathcal{N}_i$'s, at time $t_f$. During each realization
of this backward protocol a three-state transition of the type
$\Theta | \psi_m\rangle \rightarrow \Theta| a_r\rangle
\rightarrow \Theta |\psi_n\rangle$
is recorded with probability:
\begin{equation}
\widetilde P[m,r,n]=\widetilde
p(n,t_f|r,\widetilde t_1)\widetilde p(r,\widetilde t_1|m,t_0)
 \rho_{mm}
\end{equation}
where $\widetilde p(r,\widetilde t_1|m,t_0)=
|\langle a_r| \Theta^{\dagger} \widetilde U_{\widetilde
t_1,t_0}\Theta | \psi_m\rangle |^2$, and 
$\widetilde p(n,t_f|r,\widetilde t_1)=|\langle\psi_n|
\Theta^{\dagger} \widetilde U_{t_f,\widetilde t_1}\Theta  |
a_r\rangle |^2$, with $\widetilde U_{t',t}$ the time evolution
operator governed by $\widetilde H(t)$. Thus the  backward (B)
pdf of
energy and  particle number exchanges, with an interruption at
$\widetilde t_1$ is:
\begin{align} 
P_B^{\widetilde t_1}(\{ \Delta E_i\},\{\Delta N_i\})=\sum_{n,m}
\prod_i
\delta(\Delta E_i- E^i_n+  E^i_m) \nonumber \\
\times \delta(\Delta N_i- N^i_n+ N^i_m)
\widetilde p^{\widetilde t_1}(n,t_f|m,t_0) \rho_{mm}
\label{eq:PBt1}
\end{align}
where
\begin{equation}
\widetilde p^{\widetilde t_1}(n,t_f|m,t_0) = \sum_r
\widetilde p(n,t_f|r,\widetilde t_1)\widetilde
 p(r,\widetilde
t_1|m,t_0)
\end{equation}
is the conditional probability to find the system in the
state $|\psi_n\rangle$ at time $t_f$ provided that it was
in $|\psi_m\rangle$ at $t_0$ and the observable $A$ was measured
at
$\widetilde t_1$.

Using $[H(t),\Theta]=0$, and expressing the time
evolution operator as a time ordered product one finds, in a
similar way as in \cite{CTH-arXiv10}, that $\Theta^{\dagger}
\widetilde U_{\widetilde t_1,t_0}\Theta= U_{t_1,t_f}$, and
$\Theta^{\dagger} \widetilde U_{t_f,\widetilde t_1}\Theta=
U_{t_0,t_1}$. Thus $\widetilde p(n,t_f|r,\widetilde t_1)=
p(r,t_1|n,t_0)$, $\widetilde p(r,\widetilde t_1|m,t_0)=
p(m,t_f|r,t_1)$,
and consequently
\begin{equation}
p^{t_1}(m,t_f|n,t_0) = \widetilde p^{\widetilde
t_1}(n,t_f|m,t_0) 
\label{eq:pmn=tilde-pnm}
\end{equation}
follows. Combining Eqs.
(\ref{eq:rho-nn},\ref{eq:PFt1},\ref{eq:PBt1},\ref{eq:pmn=tilde-pnm})
we obtain:
\begin{align} 
\frac{P_F^{t_1}(\{ \Delta E_i\},\{\Delta N_i\})}
{P_B^{\widetilde t_1}(\{-\Delta E_i\},\{-\Delta N_i\})}
=\prod_i e^{\beta_i (\Delta E_i-\mu_i \Delta N_i)}
\label{eq:FT-t1}
\end{align}
which reads exactly as the two-measurements fluctuation theorem
\cite{Andrieux_NJP08}, with the major difference that now, a
third projective measurement is performed at $t_1$. Here the free
energy difference does not appear in Eq. (\ref{eq:FT-t1}) because
we assumed $H(t_0)=H(t_f)$. We stress that the backward and forward probabilities with intermediate measurement $P_F^{t_1}$, $P_B^{\widetilde t_1}$, in general differ from the corresponding probabilities without intermediate measurement  $P_F$, $P_B$ of Ref. \cite{Andrieux_NJP08}. Remarkably, however, their ratio remains unaltered $P_F^{t_1}/P_B^{\widetilde t_1}=P_F/P_B$.
In both cases though, particle numbers and energy measurements are assumed to be performed at  $t_0$ and $t_f$. As discussed in the introduction, these measurements are practically impossible when the subsystems are macroscopic reservoirs, as is the case with quantum transport problems. In the following we will show a way to get around this problem by taking advantage of the possibility opened by Eq. (\ref{eq:FT-t1}) of performing intermediate measurements.

The salient point in the derivation of Eq. (\ref{eq:FT-t1}) is that the generalized time reversal invariance property of the transition probability
without interruptions, i.e., $p(m,t_f|n,t_0) = \widetilde
p(n,t_f|m,t_0)$ \cite{Andrieux_NJP08}, continues to hold, Eq.
(\ref{eq:pmn=tilde-pnm}), even if a projective measurement is
performed at $t_1$. 
Evidently the result (\ref{eq:FT-t1}) can be extended to the case
of multiple intermediate quantum measurements of possibly
distinct observables $A_k$, $k=1 \dots K$ occurring at times
$t_k$ during the forward protocol, and  at times  $\widetilde t_k
=t_0+t_f-t_k$ during the backward protocol.
Thus we conclude that the
fluctuation theorem is unaffected by the action of
intermediate projective quantum measurements. 

Indeed a more detailed result holds. Consider the joint
probability for a sequence of outcomes
$n_0,\overrightarrow{n_k},n_f=n_0,n_1 \dots n_n,n_f$,
and the joint probability of the reversed sequence
$n_f,\overleftarrow{n_k},n_0 =n_f, n_n \dots n_1,n_0$
for the backward protocol:
\begin{align} 
P[n_0,\overrightarrow{n_k},n_f] &=
\prod_{k=0}^K p(n_{k+1},t_{k+1}|n_k,t_k) \rho_{n_0n_0}
\label{eq:P[trajectory]}\\
\widetilde P[n_f,\overleftarrow{n_k},n_0] &=
\prod_{k=0}^K \widetilde
p(n_k,\widetilde t_k|n_{k+1},\widetilde t_{k+1})
\rho_{n_fn_f}
\label{eq:Ptilde[trajectory]}
\end{align}
where $\widetilde t_{f,0}=t_{0,f}$, and $n_0,n_f$ label common
eigenstates of $\{H_1,H_2,\mathcal{N}_1,\mathcal{N}_2\}$, and
$n_k$ labels eigenstates of $A_k$. Using $[H(t),\Theta]=0$ one finds $\Theta \widetilde U_{\widetilde
t_{k+1},\widetilde t_k}\Theta^{\dagger}= U_{t_{k+1},t_k}$, which
leads to 
$ \widetilde p(n_k, \widetilde t_k|n_{k+1}, \widetilde t_{k+1})=
p(n_{k+1}, t_{k+1}|n_k, t_k) $.
Then, using Eqs.
(\ref{eq:rho-nn},\ref{eq:P[trajectory]},\ref{eq:Ptilde[trajectory]}),
we readily find the following detailed fluctuation theorem:
\begin{equation} 
\frac{P[n,\overrightarrow{n_k},m]}
{\widetilde P[m,\overleftarrow{n_k},n]}=
\prod_i e^{\beta_i ( E^i_m -E^i_n- \mu_i
N^i_m+\mu_i N^i_n)}
\label{eq:FT-detailed}
\end{equation}
Notably, the right hand side depends only on initial and
final eigenvalues of $H_i,\mathcal{N}_i$, but not on the
eigenvalues of $A_k$.
Eq. (\ref{eq:FT-detailed}) can be regarded as the quantum
mechanical version of a formula that was first put forward by 
Bochkov and Kuzovlev for classical trajectories (see Eq. (7) of
Ref. \cite{BochkovKuzovlev77}; a related result was derived
recently for
a classical master equation in \cite{Altland_arXiv10}).

\emph{Application to counting statistics experiments.$-$}
In the counting statistics experiment of Ref. \cite{UtsumiPRB10}
two leads having the same temperature $\beta$ are connected via a
double quantum dot. We identify the left reservoir plus left dot
as subsystem $1$, and the right reservoir plus right dot as
subsystem $2$, see Fig 1. The two subsystems are coupled through
a small (but important) time-independent interaction term $V$. At
time $t_0$ an electric potential difference $\Delta \varphi$ is
applied across the two subsystems. We model this by saying that
immediately after the time $t_0$ the density matrix is the
factorized grand canonical state $\rho$ specified above with $e
\Delta
\varphi=\mu_1-\mu_2$. Afterwards the system follows the evolution
according to $H=H_1+H_2+V$. This is equivalent to assuming that
$V$ is switched on instantaneously at $t_0$. Since $V$ is small
the energy of the total system made of the leads and the double 
quantum dot hardly changes, hence, $\Delta E_1
\simeq -\Delta E_2$. Due to the conservation of the total number of
electrons, we have $\Delta N_1=-\Delta N_2 \doteq q$.
A time-reversal invariant observable $A$, detecting whether there
is an excess electron in either of the two dots is continuously
monitored, resulting in a signal $\overrightarrow{n_k}$. The
total number of exchanged particles is calculated from
$\overrightarrow{n_k}$ by means of a function
$q[\overrightarrow{n_k}]$ that determines the number of 
transitions from the state $(1,0)$ (i.e., one electron in the
left dot and no electrons in the right dot) to the state $(0,1)$
(i.e., no electrons in the left dot and one electron in the right
dot), minus the number of opposite transitions. Evidently 
$q[\overrightarrow{n_k}]=-q[\overleftarrow{n_k}]$ holds.
The only difference with the multiple measurement scheme
presented above is that in the experiment no measurements of
total particle numbers in the two subsystems are performed at
$t_0$ and $t_f$. However since the initial density matrix is a
mixture of states with definite particle numbers in each
subsystems, a measurement at $t_0$ can be abandoned if one is
only interested in relative changes of particle numbers. By means
of registration of each transition between the two subsystems,
not only the flux of electrons is monitored but it is also
guaranteed that during the whole process the particle numbers of
each subsystem have definite values. Therefore also at $t_f$ the
state of the total system is a mixture of states with definite
particle numbers, hence a measurement of particle numbers need
not be performed at $t_f$. Thus $q[\overrightarrow{n_k}]$
coincides with the value that would have been obtained from
electron numbers measurements, i.e.,
$q[\overrightarrow{n_k}]=N^1_m-N^1_n=-N^2_m+N^2_n$. Therefore the
pdf of particle exchange is:
\begin{align} 
P_F^{\{t_k\}}(q)=\sum_{\overrightarrow{n_k}} 
\delta(q- q[\overrightarrow{n_k}])P[\overrightarrow{n_k}]
\label{eq:PF(q)}
\end{align}
where we introduced the marginal probability
$P[\overrightarrow{n_k}]=\sum_{mn}P[n,\overrightarrow{n_k},m]$.
Since the backward and forward protocols coincide in this case,
we have
$P_F^{\{t_k\}}=P_B^{\{\widetilde t_k\}}$.
Using Eqs. (\ref{eq:FT-detailed},\ref{eq:PF(q)}) with
$q[\overrightarrow{n_k}]=-q[\overleftarrow{n_k}]$, we conclude
that the pdf of electron exchange indeed satisfies:
\begin{equation} 
P_F^{\{t_k\}}(q)= P_F^{\{t_k\}}(-q)e^{\beta(\mu_1-\mu_2)q}
\label{eq:FT-ex}
\end{equation}
in good qualitative agreement for small small $\tau$, and excellent quantitative agreement for large $\tau$, 
 with experimental findings of Ref. \cite{UtsumiPRB10}, see Fig. 2a therein \cite{noteX}. The same experimental setup
could be used to confirm the validity of the detailed fluctuation
relation 
\begin{equation}
 P[\overrightarrow{n_k}]=P[\overleftarrow{n_k}]
e^{\beta(\mu_1-\mu_2)q[\overrightarrow{n_k}]}
\label{eq:FT-ex-detailed}
\end{equation}
that follows from (\ref{eq:FT-detailed}).

\emph{Remarks.$-$}
One assumption of our theory is that the initial state of the system is
the direct product  of two grand-canonical states. For this
particular choice of the initial state the fluctuation theorems
(\ref{eq:FT-ex},\ref{eq:FT-ex-detailed}) do hold
independently of the duration $\tau$ of the protocol. In
experiments, however,
unavoidable interactions may lead to correlations between the
sub-parts of the system already before the protocol has
started. 
These initial correlations in general will lead to transient
deviations from the fluctuation theorems which then are expected
to hold as steady state fluctuation theorems in the limit of
large $\tau$, as is the case with experiment \cite{UtsumiPRB10}.

The assumptions that the measured observables $A_k$ have
non-degenerate eigenvalues can be abandoned without any change of
our central result formulated in Eqs. (\ref{eq:FT-t1},
\ref{eq:FT-detailed}).  We further found that our main result
remains true if the quantum measurements are performed with
respect to
positive operator valued measures (POVM's)
instead of von Neumann projection valued measures. This is not so
surprising after all, being the former less invasive than the
latter.
The main result holds for any number of subsystems.
In particular for a single closed system it generalizes the Tasaki-Crooks work
fluctuation theorem \cite{Tasaki00,TLH_PRE07,TH_JPA07}. In case of two susbsystems with equal temperature and no matter exchange it generalizes the Tasaki-Crooks fluctuation theorem for open systems
\cite{TCH_JStatMech09,CTH-PRL09}. Multiple measurements of reservoir energies (not fluxes) were considered for this case in Ref. \cite{DeRoeck04PRE69}.

Far from being a problem of purely academic interest, the
multiple measurement scheme pursued here, helps bridging
the
existing gap between theory and experiments concerning quantum
fluctuation theorems.
The measurement of energy and particle content of the subparts of
a system may be practically impossible, especially when the
subparts are macroscopic objects like the leads in an electron
counting statistics experiment.
On the other hand energy and matter exchanges may be measured by
\emph{monitoring the respective fluxes through interfaces}, making the
cumbersome measurements of total energies and particle numbers
obsolete. Here we have shown that the fluctuation theorem
continues to hold in this case of continuous monitoring, Eqs.
(\ref{eq:FT-t1},\ref{eq:FT-detailed}), provided a theoretical explanation for an
experimentally observed relation, Eq. (\ref{eq:FT-ex}), and
predicted that a more detailed relation, Eq.
(\ref{eq:FT-ex-detailed}) should hold in the same set-up.

This work was supported by the cluster of excellence
Nanosystems Initiative Munich (NIM) and the Volkswagen Foundation
(project I/83902).


\begin{thebibliography}{20}
\bibitem{noteX}
In Ref. \cite{UtsumiPRB10} the measurement apparatus is not only
responsible for wave function collapses but also exchanges
electrons. According to \cite{UtsumiPRB10}, this exchange of
electrons, that was not considered here, leads to a
renormalization of the temperature in (\ref{eq:FT-ex}).
\end{thebibliography}
\end{document}